# Deep Impact: Optical Spectroscopy and Photometry Obtained at MIRA

Russell G. Walker, Wm. Bruce Weaver, W. W. Shane, and Arthur Babcock

Monterey Institute for Research in Astronomy (MIRA), 200 Eighth Street,

Marina, California, 93933

Number of pages = 43

Number of figures = 16

Number of Tables = 3

# ABSTRACT

We present spectroscopic and photometric observations, spanning the optical UV to the far red, before, during, and after the NASA Deep Impact event of July 4, 2005. The inner 2000 km of the pre and post-impact coma was about 0.3 magnitude redder in B-R than in the outer coma. The pre-impact spectrum was a faint reflected solar spectrum dominated by molecular emissions extending > 40000 km from the nucleus. The post-impact light curve in R and I showed a rapid rise consistent with an expanding optically thick cloud during the first 18 minutes after impact. During the next 8 minutes the cloud became optically thin. Sixty minutes after impact the impact R-band flux reached a plateau at $7.5 \times 10^{-15}$ erg cm$^{-2}$ s$^{-1}$ Å$^{-1}$, the comet brightening by a factor of ~4.3 above its pre-impact value observed in a 15″ aperture. The mean expansion velocity of the grains during the first 49 minutes was $229 \pm 49$ m s$^{-1}$. The spectrum became dominated by scattered sunlight during the first hour after impact. The volume scattering function (VSF) observed 32 minutes after impact shows strong reddening. At 49 minutes, however, the VSF shows an additional two-fold increase in the blue but only a 20 per-cent increase at 5500Å. Post-impact spectra and R-I photometry showed rapid reddening. The particle size distribution, dominated by 1 to 2.5 micron particles shortly after impact, changed dramatically during the first hour due to sublimation of water-ice particles of this size. On the night following impact the comet was still substantially brighter than before impact, but R-I had returned to its pre-impact value. B-R remained significantly redder. The ejecta 25 hours after impact was fan-shaped subtending ~180° roughly symmetrical about position angle 225°. The mean expansion velocity 90° from the direction to the Sun was $185 \pm 12$ m s$^{-1}$.

# 1. OBSERVATIONS

Spectroscopic and imaging observations of Comet 9P/ Tempel 1 were obtained on four nights starting on June 29, 2005 UT (Table 1) with the 36-inch telescope at MIRA's Bernard Oliver Observing Station (OOS). As the telescope Guidance Acquisition Package can rapidly divert light between instruments, both CCD imaging and spectroscopic observations were made. The nights were photometric and the seeing was typically ~2″ for the range of air masses (1.6 to 3.1) at the times of observation. Our goal was to detect and record changes in the morphology and the photometric and spectroscopic signatures of the comet that were produced by the Deep Impact (DI) event.

Table 1 is a chronological list of our observations. Pre-impact measurements were obtained on June 29th, July 3rd, and early on July 4th UT. Post-impact observations were obtained on July 4th and July 5th.

Table 1.

## 1.1 Spectroscopy

Spectroscopic observations were made using an Andor camera with a back-illuminated e2v 1024 x 512 CCD array antireflection-coated for the optical UV and cooled to -85 C. The log of spectroscopic exposures is included in Table 1. The dispersion of the spectra was 2.05 Å pixel$^{-1}$ and the slit width was set to 3 pixels (1.8″) resulting in a measured FWHM of 7Å. A 79″ long slit was placed across the coma in the east-west direction. The wavelength coverage was 3341–5466Å. After reducing the CCD spectral images in the usual way for flat and bias, a sky spectrum was subtracted. The wavelength sensitivity of the system was determined by observations of α Lyr and the secondary standard μ Her, G5 IV (Berger, 1976). Mu Her was normalized to the irradiance of the solar spectrum (Neckels, 1999) and used as the solar reference spectrum in our analysis. Checks of the photometric calibration between α Lyr and μ Her show errors of a few percent except below 3400Å, where they rise to about 15 percent.

Depending on the quality of the guiding, spectra of the inner coma were made by integrating the response in the central 12″ to 27″ of the long-slit spectra; the typical width was 15″. This length corresponds to ~10,000 km at the comet. This was done to minimize the substantial contribution of the extended coma and to focus on just the central part of the coma. The spectra were corrected for the effects of airmass with the standard atmospheric model for the OOS, converted to energy flux units, and corrected for the wavelength sensitivity of the instrument. The data were lightly smoothed with a flux-conserving five-point Gaussian filter. Because of the extended nature of the coma, the effects of differential refraction along the slit were negligible.

## 1.2 Imaging Photometry

Imaging observations were made with a Fairchild Imaging Peregrine-447 camera, with a 2048 x 2048 back-illuminated array of 15μm pixels, yielding a plate scale of 0.34 arcsec pixel$^{-1}$. Observations were made primarily in Kron-Cousins R and I, with a few observations in B. The objective was to monitor the solar radiation scattered from solid material in the cloud of impact ejecta. The R and I bands were selected because gas emission was expected to be only a minor component to the flux at these wavelengths.

Calibration files consisting of bias, dark, and twilight sky flat fields were made on each night. Initial image reduction was accomplished by subtracting bias and dark frames and dividing by flat field frames using Mira Pro software.[1]

We established sets of secondary standard stars, 12 in each field in which the comet was observed. Field centers are given in the Appendix. This was done on February 23, 2006, using the same instrumental arrangement and reduction method as was used for the comet observations. On the same night we observed 65 stars in Selected Areas 98, 101, and 104 (see Blaauw and Elvius, 1965). The regions observed were determined by the field of our camera and were selected so as to include a sufficient number of stars with well determined magnitudes and with as large as possible a range of colors. For these stars we adopted the magnitudes provided by Stetson (2000). We determined magnitudes for our secondary standards (also included in the Appendix) using the IRAF[2] photometric reduction package and the curve of growth method described by Stetson (1990). Then using the same methods, with slight modifications because of the nature of the comet image, we determined magnitudes for the comet from each observation and within 11 apertures varying from 2.7″ to 30.6″ diameter. The mean internal error of a single magnitude determination, as reported by the IRAF software, is about 0.005 in most cases. The scatter amongst redundant observations indicates an error closer to 0.01 magnitude, with occasional higher values. Short term variation in the brightness, particularly after impact, appears to have contributed to these errors.

---

[1] Mira Pro software is published by Mirametrics, Inc., which has no connection with the Monterey Institute for Research in Astronomy.

[2] IRAF is distributed by the National Optical Astronomy Observatories, which are operated by the Association of Universities for Research in Astronomy, Inc., under cooperative agreement with the National Science Foundation.

Using the curve of growth method for these observations introduces several complications. It is intended, when used for stellar photometry, to correct for the loss of light to regions outside the measuring aperture due to seeing and other effects. We necessarily determined the curve of growth for each observation from the stellar images. In doing so, we corrected for the bright nucleus of the comet, but we undercorrected for those parts of the comet image lying within but near to the edge of the aperture and we did not correct for light scattered into the aperture from those parts of the comet image lying outside but near the edge of the aperture. These two errors work in opposite directions, but they do not necessarily compensate. A further complication is introduced by the comet motion. We tracked the comet during the exposures, so that the stellar images are trailed over a distance of 2.4″ during a 100 second exposure. This is slightly less than the 2.7″ diameter of the smallest aperture, and it may be expected to result in an overcorrection by the curve of growth method. This may account for the slight upward curvature at the smallest aperture in the flux vs. aperture diagrams.

## 2. RESULTS

### 2.1 Spectroscopy

A contour plot of the spectral image recorded just prior (5:34 UT, July 4, 2005) to the DI event (05:52:02 UT ) is shown in Fig. 1. The spectrum is dominated by emission bands from familiar molecular species: CN (0,0) at 3883 Å that extends more than 40,000 km from the nucleus; C3 at 4050 Å; the C2 ($\Delta v=0$), C2 ($\Delta v=1$), C2 ($\Delta v=2$) Swan bands, and fainter bands of CN and $NH_2$. The dust continuum is much less evident.

This may be compared to the continuum in the spectrum taken 49 minutes after impact shown in Fig.2, where scattered sunlight from the dust clearly dominates.

Fig. 1 and Fig 2

The continuum profile (orthogonal to the wavelength scale) from 4500 Å to 5400 Å was used to estimate the mean velocity of the dust grains during the first 49 minutes after impact to be 229 ± 49 m/s, the uncertainty being largely defined by guiding errors. This value is somewhat higher than the 130 m/s measured by Jehin, et al. 2006, and the 160 m/s measured by the Rosetta Spacecraft (Keller, et al. 2006), but comparable to the 220 m/s observation of Schleicher, et al. 2006.

All of the spectra taken at MIRA are plotted for comparison in Fig. 3. The four pre-impact spectra all cluster in the lower part of the graph along with the post-impact spectrum of July 5$^{th}$, indicating that the impact dust had largely dissipated and that the coma of 9P/Tempel 1, both dust and gas, had recovered to near pre-impact values. Small apparent excesses seen in the Swan bands are consistent with that expected from rotation of the nucleus (see Jehin, et al. 2006). The exception is the CN intensity which still shows an excess of ~1.7 over the pre-impact value (also seen in Fig.4). There is an increase in solar features such as Ca II H & K, the G-band, Hβ, the Mg I complex at λ5170, and the Ca I – Fe I complex at λ5270 in the post-impact spectra on July 4th proportional to the increased continuum radiance. These features also return to near pre-impact levels on the 5 July spectrum.

Fig. 3

The post-impact spectra of July 4[th] (6:24 and 6:41 UT) show a strong and rapid increase of the dust component and a mild increase in CN emission. The C2 and C3 bands appear to be little affected by the impact at this time. We have subtracted the pre-impact spectra from the post-impact spectra and divided the result by the solar irradiance (µ Her) incident on the comet. This procedure yields a measure of the wavelength dependence of the difference in the line-of-sight integral of the volume scattering function, VSF (Grun and Jessberger, 1990), of the post-impact grains from that of the pre-impact grains in spectral regions where the molecular emissions have not changed. The resulting curves for spectra taken 32 and 49 minutes after impact are plotted in Fig. 4. The mean of the two spectra taken at 05:19and 05:34 UT on July, 4[th] was adopted as the pre-impact spectrum.

The two curves show the strong increase in the flux of the CN band and the onsets of the smaller increases in the flux of C3 and the Swan bands. The spectral shape of the scattered dust continuum has evolved rapidly. At impact plus 32 minutes it has become strongly reddened relative to its pre-impact shape. Seventeen minutes later, while the differences continue to grow, there is a preference for growth in the blue over that in the red. The VSF is dependent upon the size distribution and scattering efficiencies (related to composition and shape) of the entire ensemble of particles within the observing aperture. The observed change in the VSF implies a large change in the observed particle size distribution. Detailed modeling of the spectral behavior of these data, which is beyond the scope of the present paper, will constrain the possible mechanisms. However, if the ejecta cloud has become optically thin by this time (see R-band photometry discussed below), then the increase in the blue is likely due to scattering by a population of small particles released from larger sublimating ice grains.

Fig. 4

Fig. 5

Fig. 5 shows the changes observed in the CN(0,0) band during the DI event. The CN data have been normalized to the R-band photometry (discussed below) just prior to impact. It is clear that production of CN due to the impact lags production of the dust. This is consistent with the long dissociation lifetime (~50,000 s) determined by the Rosetta spacecraft (Keller, et al. 2005) for the parent molecule. The CN flux that we see 49 minutes after impact is ~25% greater than that predicted by the CN gas emission rotational phase curve (Jehin, et al. 2006) with a period of 1.709 days based on their measurements prior to impact.

## 2.2 Images

Fig. 6

The images in Fig. 6 show the evolution of the DI dust ejecta from 2 minutes before impact until 25 hours after impact. The three early frames are centered on the nucleus of 9P/Tempel 1 and show the brightening of the coma and its elongation in the direction of plume ejection at position angle ~225º (Meech et al. 2005). The fourth image is a co-addition of 6 images taken at a mean time of 25 hours after impact. We have subtracted the image taken 2 minutes before impact from the co-added image to enable the diffuse dust fan to be seen. At this time the dust has moved 15120 km in the 225º direction with a mean velocity of 177 ± 12 m/s as projected on the sky. The mean velocity since impact has been determined from measures of the angular distance of a point on the faintest contours (2 ADU, 3ADU) to the nuclear position of the comet. These measurements are plotted in Fig. 7 for the range of position angles (150º to 330º) subtended by the evolved impact fan. The slope of the data is a consequence of solar radiation pressure on grains emitted into a fairly wide cone, and the cloud geometry.

Initially the grains may have been ejected from the nucleus into a cone of, say, 90º, but sublimation of ice will release additional small grains isotropically from centers of mass that are coupled with the outflow velocity effectively widening the fan. Solar radiation pressure will preferentially push the small grains away from the Sun, which is at a position angle of 290º. The dust cloud will not progress as far (in a given time) from the nucleus in the solar direction where the radiation pressure decelerates the grain velocity as in the anti-solar direction where it accelerates the grains. The deceleration due to radiation pressure was estimated to be $0.0389 \pm 0.004$ cm s$^{-2}$ from the difference in velocities observed in the directions of the Sun and 90º from it. The solar gravitational acceleration at the comet was 0.26102 cm s$^{-2}$ leading to a radial radiation pressure constant, $\beta = F_{rad}/F_{grav} = 0.149$. For spherical homogeneous particles $\beta = 1.2 \times 10^{-4} Q_{pr}/(\rho d)$ (Krishna Swamy, 1986). If the efficiency factor for radiation pressure, $Q_{pr} = 1$ and taking for silicates a bulk density, $\rho = 2.37$ gm cm$^{-3}$ (Lamy, 1974), yields a grain diameter, $d = 3.4$ µ for the particles at the edge of the dust fan. This result is based on a single spherical particle model. The problem becomes more complicated for if the grains are irregular in shape, fluffy, porous, or aggregates of very small grains. In the case of fluffy aggregates (Kimura, et al., 2002) the radiation pressure constant will have a nonradial component in addition to the radial one due to the asymmetry of the scattering. This leads to $\beta$ values for aggregates of sub-micron particles that are less than those of spherical grains with the same volume-equivalent diameters, but which approach the spherical values as the aggregate mass decreases. Thus we may consider 3.4 µ an upper limit to the grain size for silicates at the edge of the fan.

Fig. 7

## 2.3 Photometry

Figures 8 through 13 present a summary of the comet imaging photometry. Except as noted in the figure captions, these are based on means of two to six observations made in succession. In these figures, apertures have been converted to the diameter of corresponding cylinders at the distance of the comet and brightness is expressed in units of flux (Drilling and Landolt, 2000). The remaining figures show the dependence upon aperture of color index or total flux within the aperture as determined on each night of observation.

The curves plotted in Fig. 8 clearly show that in all three bands the post-impact fluxes observed on July 5$^{th}$ have not recovered to the pre-impact values measured on June 29$^{th}$. All the curves are quite linear beyond ~5000 km from the nucleus, confirming the expected r$^{-1}$ brightness distribution within the coma.

Figure 9 shows the B-R and R-I colors derived from the fluxes plotted in Fig. 8 pre- and post-impact. With the exception of a possible bluing in the inner coma on July 29$^{th}$, the R-I colors observed both nights are essentially the same and invariant (R-I ≈ 0.43) with distance from the nucleus even though R and I (Fig. 8) have not returned to their pre-impact state. In contrast, the B-R color remained significantly redder after impact with the pre-impact color becoming bluer with increasing distance from the nucleus. Both pre- and post-impact data are bluer than the solar value (B-R = 1.19) for distances greater than 6000 km. This is a consequence of the B-filter including flux from the CN(0,0) band.

In determining the color index vs. aperture relations in Fig. 9, we measured the fluxes within each annulus, as defined by the successive apertures, and converted this flux to a magnitude. The first annulus had zero inner radius. The second annulus, between 879 and 1538 km, covered a small area and contained substantially less flux than any of the others, accounting for the uncommonly large uncertainties. In Fig. 9 the B-R curves depend upon only one or two B observations, and the color spikes at 1200 km in both curves are of about the amplitude of the uncertainties. The spike in one of the R-I curves is somewhat larger than the corresponding uncertainty, but it depends largely on a single deviant observation. The color reversal for the smallest aperture suggests that application of the curve of growth may have been responsible for this deviation.

Fig. 8 and Fig. 9

Observations on July 3$^{rd}$ were made in R only. These are plotted in Fig. 10 for the same set of apertures as in Fig. 8. There was little change in the flux during the 2 hours between observations. The $r^{-1}$ brightness profile works well into the inner coma. The flattening of the curve at the faintest two apertures is due to overcompensation of the images of stars that were trailed while the comet image was not trailed.

Fig 10 and Fig 11

R and I measurements made about an hour before and an hour after impact are shown in Fig. 11. To a very good first order the pre-impact flux curves have just been displaced upward due to the flux from the impact cloud which, even at +75 minutes, is still totally contained within the minimum aperture diameter.

R-I colors 69 minutes after impact are compared in Fig.12 to those 54 minutes prior to impact. These data were reduced in the same manner as those of Fig. 9. The color spike at 1200 km in the post-impact observations suggests that the impact cloud is extremely red. The uncertainties in measurements at this radius, as discussed in connection with Fig. 9, are equally applicable here. However, in this case the amplitude of the color spike is almost twice the calculated uncertainty, so we are inclined to credit its reality. An additional uncertainty arises from the non-simultaneity of the R and I observations, the I being slightly later than the R. This would suggest that more material was in the second annulus during the I observations than during the R.

Fig. 12

If we are prepared to accept the large R-I color in the 1200 km annulus, then this limits the size of the scattering particles. For example, Fig. 13 shows the R-I resulting from Mie[3] scattering of sunlight from water ice spheres at the scattering angle of 139 degrees. For this simple model, the dominant particle radii are limited to $1 - 2.5\mu$. Figure 4 shows that, by 49 minutes after impact, the spectrum is becoming bluer and closer to the spectral distribution of sunlight, suggesting a loss of the $1 - 2.5\mu$ ice particles. These results are consistent with the model described by Keller et al. (2005) in which more water was observed than could be sublimed by the energy of the impactor by three orders of magnitude. While more complicated particle shapes are not unlikely, adding more free parameters to this analysis is beyond the scope of this paper. A similar Mie scattering analysis for silicate spheres requires particle radii of the order of 0.5µ, consistent with the results of Harker et al. (2005); however, they do not see a decline in their silicate features until 1.8 hours after impact. The most likely case is that both particle types are present but the relatively rapid decline in reddening indicates that the water ice is the primary optical component.

---

[3] Mie scattering calculations were performed with MieTab 8.33.4, created by August Miller of New Mexico State University and WJ Lentz of the Naval Postgraduate School, available at http://amiller.nmsu.edu

Fig 13

The flux history of the DI event is shown by the light curves in Fig.14. Here we have pre-impact baselines, the increase of the flux from the expanding cloud, and an apparent plateau at the end of our night's observing. Gaps in the curves are the times when spectroscopic observations were made. The R and I observations have been combined by shifting the less plentiful I-magnitudes by the R-I values determined before or well after impact, as appropriate. For the observations made shortly after impact we used the mean of the before and after values. The color change, as shown in Fig.12, is sufficiently small for apertures greater than 2500 km that this approximation seems justified.

Fig 14

Figure 15 was constructed by subtracting the flux observed 2 minutes before impact from the flux in each aperture shown in Fig.14. The light curve in Fig. 15 has several distinct regions. There is a steep rise in the flux during the first 2 minutes after impact followed by a concave portion of lesser slope from 2 to 18 minutes. The flux then rolls over from 18 to 25 minutes to a region of still lesser slope, and finally a gentle flattening of the curve 62 to 76 minutes after impact with a possible flux maximum at about 71 minutes after impact. A simple optically thick sphere expanding at constant velocity successfully models the data from the concave portion of the light curve. In this case the flux incre ases with the square of the radius (velocity x time), and a plot of the square root of the flux versus time should be linear. Although models are seldom unique and the ejecta cloud observed by the spacecraft was not spherical, we believe that Fig.16 provides strong support for the contention that the cloud of ejecta was optically thick during this time period.

Fig. 15 and Fig 16

### 3.0 Summary and Conclusions

Calibrated CCD photometric and long-slit spectrographic observations of the Deep Impact event were made at MIRA. Taken together, these observations spanned a range from the optical UV to the far red.

Pre-impact observations confirmed the expected $r^{-1}$ brightness distribution within the coma. The inner 2000 km of the coma was about 0.3 magnitude redder in B-R, both before and one day after impact. The pre-impact spectrum was dominated by molecular emission extending beyond 40000 km from the nucleus with a faint reflected solar spectrum.

No impact flash was detected. The post-impact light curve in R and I showed a rapid rise consistent with an expanding optically thick cloud. In the first 18 minutes after impact the R-band flux increased by 5 x $10^{-15}$ erg $cm^{-2}$ $s^{-1}$ $Å^{-1}$. During the following 8 minutes the cloud was clearly becoming optically thin. Sixty minutes after impact the R-band flux due to the impact reached a plateau at 7.5 x $10^{-15}$ erg cm$^{-2}$ s$^{-1}$ Å$^{-1}$, the comet brightening by a factor of ~4.3 above its pre-impact value observed in a 15″ aperture.

During the first hour after impact the spectrum was dominated by scattered sunlight. The observed volume scattering function 32 minutes after impact shows a strong reddening. At 49 minutes, however, it shows a two fold increase in the blue but only a 20 percent increase at 5500 Angstroms. The estimated expansion velocity of the dust grains during the first 49 minutes was 229 ± 49 m s$^{-1}$. The 49-minute lag in the increase in the CN(0,0) band is consistent with lifetime of the parent molecules.

The post-impact spectra and the R-I photometry show rapid reddening. The particle size distribution, dominated by 1 to 2.5 micron particles shortly after impact, changes dramatically during the first hour. This can be explained by the prompt sublimation of water-ice particles of this size.

On the night following impact the comet was still substantially brighter than before impact, but the R-I color had returned to its pre-impact value. However, the B-R color remained significantly redder. The ejecta 25 hours after impact appeared as a fan subtending 180 degrees roughly symmetrical about position angle 225. The expansion velocity at a position angle 90 degrees from the direction to the Sun was 185 ± 12 m s$^{-1}$.

## Appendix - Photometry of Secondary Standards

Table A-1 is a list of the centers of each field observed for secondary standards, along with the date on which Comet 9P/ Tempel 1 was observed in that field. Table A-2

shows the coordinates of each of our 48 secondary standard stars, with magnitudes and an estimate of the magnitude error for each filter in which that field was observed.

Table A-1

Table A-2

## Acknowledgements

The authors gratefully acknowledge the support of Fairchild Imaging, the Ralph Knox Foundation, the Kenneth Lafferty Hess Family Charitable Foundation, and Mirametrics, Inc. We thank Holly Keifer, Jim Neeland, and Karl Gramespacher for their technical work on this project. A.B. and W.S. are guest users of the Canadian Astronomy Data Center, which is operated by the Dominion Astrophysical Observatory for the National Research Council of Canada's Herzberg Institute of Astrophysics. The Monterey Institute for Research in Astronomy owns and operates the Oliver Observing Station under permit from the U. S. Department of Agriculture-Forest Service, and owns and operates the Richard W. Hamming Astronomy Center and the Ralph Knox Shops through an arrangement with the U.S. Department of Education.

Quantities, 4th Edition, Springer-Verlag New York, Inc. pp. 353-354.

Schleicher, D. et al, 2006. Photometry and Imaging Results for Comet 9P/Tempel 1 and *Deep Impact*: Gas Production Rates, Postimpact Light Curves, and Ejecta Plume Morphology. A.J. 131, pp. 1130-1137.

Stetson, P. B., 1990. On the Growth-Curve Method for Calibrating Stellar Photometry With CCDs. PASP 102, pp. 932-948.

Stetson, P. B., 2000. Homogeneous Photometry for Star Clusters and Resolved Galaxies. II Photometric Standard Stars. PASP 112, pp. 925-931.

## Table 1. Observing Log

| Pre-impact Observations | | | | Post-impact Observations | | | |
|---|---|---|---|---|---|---|---|
| UT Date/Time (2005) mm/dd-hh:mm | Exposure (seconds) | Filter or Spectrum | Comet Magnitude | UT Date/Time (2005) mm/dd-hh:mm | Exposure (seconds) | Filter or Spectrum | Comet Magnitude |
| 06/29-05:25 | 30 | R | 13.50 | 07/04-05:54 | 100 | R | 13.45 |
| 06/29-05:28 | 30 | R | 13.46 | 07/04-05:57 | 100 | R | 13.38 |
| 06/29-05:35 | 100 | R | 13.49 | 07/04-06:03 | 100 | I | 12.74 |
| 06/29-05:37 | 100 | R | 13.48 | 07/04-06:06 | 100 | I | 12.65 |
| 06/29-05:39 | 100 | R | 13.48 | 07/04-06:08 | 100 | R | 13.00 |
| 06/29-05:41 | 100 | R | 13.49 | 07/04-06:10 | 100 | R | 12.95 |
| 06/29-05:44 | 100 | R | 13.46 | 07/04-06:13 | 100 | R | 12.90 |
| 06/29-05:47 | 100 | R | 13.48 | 07/04-06:15 | 100 | R | 12.87 |
| 06/29-05:50 | 100 | I | 13.03 | 07/04-06:17 | 20 | R | 12.91 |
| 06/29-05:53 | 100 | I | 13.04 | 07/04-06:24 | 900 | spectrum | 15″ |
| 06/29-05:56 | 100 | I | 13.04 | 07/04-06:41 | 900 | spectrum | 16″ |
| 06/29-05:58 | 100 | I | 13.07 | 07/04-06:54 | 100 | R | 12.75 |
| 06/29-06:00 | 100 | I | 13.05 | 07/04-06:56 | 100 | R | 12.75 |
| 06/29-06:02 | 100 | I | 13.04 | 07/04-07:01 | 100 | R | 12.74 |
| 06/29-06:05 | 100 | B | 14.66 | 07/04-07:03 | 100 | R | 12.74 |
| 06/29-06:36 | 600 | Spectrum | - | 07/04-07:06 | 100 | I | 12.29 |
| | | | | 07/04-07:08 | 100 | I | 12.30 |
| 07/03-04:53 | 100 | R | 13.57 | | | | |
| 07/03-04:55 | 100 | R | 13.56 | 07/05-04:54 | 100 | I | 12.82 |
| 07/03-04:57 | 100 | R | 13.57 | 07/05-04:57 | 100 | I | 12.83 |
| 07/03-04:59 | 100 | R | 13.57 | 07/05-05:00 | 100 | I | 12.81 |
| 07/03-05:01 | 100 | R | 13.57 | 07/05-05:02 | 100 | R | 13.23 |
| 07/03-05:03 | 100 | R | 13.56 | 07/05-05:04 | 100 | R | 13.25 |
| 07/03-05:19 | 300 | Spectrum | ⎫ | 07/05-05:07 | 100 | R | 13.25 |
| 07/03-05:40 | 1800 | Spectrum | ⎬ 27″ | 07/05-05:09 | 100 | R | 13.24 |
| 07/03-06:37 | 900 | Spectrum | ⎭ | 07/05-05:11 | 100 | R | 13.24 |
| 07/03-06:51 | 100 | R | 13.57 | 07/05-05:14 | 100 | R | 13.25 |
| 07/03-06:54 | 100 | R | 13.60 | 07/05-05:16 | 100 | B | 14.49 |
| 07/03-06:56 | 100 | R | 13.58 | 07/05-05:18 | 100 | B | 14.50 |
| 07/03-06:59 | 100 | R | 13.58 | 07/05-05:22 | 30 | I | 12.82 |
| 07/03-07:01 | 100 | R | 13.58 | 07/05-05:24 | 30 | I | 12.82 |
| 07/03-07:03 | 100 | R | 13.57 | 07/05-05:26 | 30 | I | 12.84 |
| | | | | 07/05-05:33 | 300 | spectrum | ⎫ |
| 07/04-04:50 | 100 | R | 13.60 | 07/05-05:51 | 1800 | spectrum | ⎬ 21″ |
| 07/04-04:52 | 100 | R | 13.60 | 07/05-06:22 | 1800 | spectrum | ⎭ |
| 07/04-04:55 | 100 | R | 13.59 | 07/05-06:42 | 100 | R | 13.27 |
| 07/04-04:58 | 100 | R | 13.60 | 07/05-06:44 | 100 | R | 13.27 |
| 07/04-05:00 | 100 | I | 13.17 | 07/05-06:47 | 100 | R | 13.28 |
| 07/04-05:03 | 100 | I | 13.13 | 07/05-06:49 | 100 | R | 13.27 |
| 07/04-05:05 | 100 | I | 13.16 | 07/05-06:52 | 100 | R | 13.28 |
| 07/04-05:18 | 600 | Spectrum | 17″ | 07/05-06:54 | 100 | R | 13.27 |
| 07/04-05:34 | 900 | Spectrum | 12″ | 07/05-06:57 | 100 | I | 12.84 |
| 07/04-05:50 | 100 | R | 13.60 | 07/05-06:59 | 100 | I | 12.84 |
| 07/04-05:52 | 100 | R | 13.56 | 07/05-07:02 | 100 | I | 12.84 |
| 07/04-05:52 | | IMPACT | | 07/05-07:04 | 100 | I | 12.84 |

Notes: The listed UT dates/times are for the midpoint of the exposures. The comet magnitudes were measured in a 17.0 arcsec diameter circular aperture positioned at the centroid of the comet radiance. The spectral width in arcsec used for extracting the inner coma spectra is given in the magnitude column for the spectral observations.

Table A-1. Field centers of the 4 fields observed for the purpose of establishing secondary standard stars with the UT date when the comet was observed in each field.

| Field | α(2000.0) | δ(2000.0) | Comet in Field |
|---|---|---|---|
| 1 | 13 29 13.0 | -07 36 04 | June 29 2005 |
| 2 | 13 36 13.2 | -09 12 23 | July 3 2005 |
| 3 | 13 37 56.2 | -09 35 31 | July 4 2005 |
| 4 | 13 39 47.5 | -09 58 47 | July 5 2005 |

Table A-2. Coordinates (for identification only) and magnitudes of the 48 secondary standard stars as observed February 23, 2006. Error estimates (1 σ) are based on IRAF estimates of internal error and scatter in redundant observations. Stars that showed high residuals (> .05 mag) when these magnitudes were used to determine photometric zero points for the comet observations of July 2005 were excluded from the fit as possible variables. They are identified in the Notes column.

| Field | α(2000.0) | δ (2000.0) | B | σ(B) | R | σ(R) | I | σ(I) | Notes |
|---|---|---|---|---|---|---|---|---|---|
| 1 | 13 28 51.6 | -07 35 21 | 16.635 | .018 | 15.432 | .008 | 15.018 | .009 | |
| 1 | 13 28 54.9 | -07 33 00 | 17.975 | .018 | 16.640 | .013 | 16.218 | .014 | |
| 1 | 13 28 54.9 | -07 41 22 | 16.318 | .011 | 15.265 | .005 | 14.900 | .006 | |
| 1 | 13 28 55.3 | -07 36 40 | 14.138 | .002 | 13.266 | .002 | 12.974 | .003 | |
| 1 | 13 29 04.0 | -07 36 10 | 16.824 | .008 | 14.556 | .005 | 13.799 | .003 | |
| 1 | 13 29 06.6 | -07 40 25 | 15.925 | .005 | 14.715 | .004 | 14.345 | .007 | |
| 1 | 13 29 07.3 | -07 35 40 | 14.294 | .002 | 13.238 | .002 | 12.904 | .003 | |
| 1 | 13 29 10.9 | -07 33 50 | 17.060 | .014 | 15.655 | .007 | 15.168 | .010 | |
| 1 | 13 29 14.4 | -07 34 38 | 14.639 | .003 | 13.725 | .003 | 13.418 | .004 | |
| 1 | 13 29 20.8 | -07 37 45 | 17.171 | .011 | 15.541 | .005 | 15.043 | .009 | |
| 1 | 13 29 21.4 | -07 38 13 | 17.218 | .012 | 15.932 | .009 | 15.510 | .012 | |
| 1 | 13 29 29.5 | -07 40 52 | 15.189 | .008 | 13.927 | .006 | 13.539 | .008 | |
| 2 | 13 35 51.1 | -09 12 02 | | | 17.169 | .015 | 15.796 | .0 | variable? |
| 2 | 13 35 52.6 | -09 07 31 | | | 15.926 | .014 | 15.506 | .014 | variable? |
| 2 | 13 35 55.5 | -09 05 59 | | | 12.632 | .002 | 12.303 | .002 | |
| 2 | 13 35 57.2 | -09 15 07 | | | 16.039 | .011 | 15.463 | .015 | |
| 2 | 13 36 02.5 | -09 12 28 | | | 14.807 | .008 | 14.373 | .011 | |
| 2 | 13 36 04.0 | -09 07 39 | | | 15.272 | .008 | 14.715 | .008 | |
| 2 | 13 36 09.5 | -09 07 50 | | | 14.746 | .004 | 14.402 | .005 | |
| 2 | 13 36 11.8 | -09 11 42 | | | 16.576 | .012 | 16.232 | .016 | |
| 2 | 13 36 11.9 | -09 12 57 | | | 14.231 | .010 | 13.914 | .013 | |
| 2 | 13 36 13.6 | -09 10 03 | | | 16.291 | .010 | 15.699 | .013 | |
| 2 | 13 36 18.6 | -09 12 21 | | | 15.677 | .010 | 15.246 | .014 | |
| 2 | 13 36 25.1 | -09 07 24 | | | 15.414 | .007 | 14.784 | .006 | |
| | | | | | | | | | |
| 3 | 13 37 37.7 | -09 33 41 | | | 16.176 | .011 | 15.881 | .011 | variable? |
| 3 | 13 37 39.7 | -09 37 00 | | | 15.881 | .008 | 15.426 | .007 | |
| 3 | 13 37 42.5 | -09 34 55 | | | 15.432 | .010 | 15.033 | .007 | |
| 3 | 13 37 49.2 | -09 34 26 | | | 15.636 | .009 | 15.040 | .007 | |
| 3 | 13 37 50.1 | -09 40 13 | | | 15.566 | .008 | 15.144 | .006 | |
| 3 | 13 37 51.4 | -09 36 00 | | | 14.296 | .004 | 13.915 | .005 | |
| 3 | 13 37 55.8 | -09 35 41 | | | 16.651 | .017 | 15.264 | .007 | |
| 3 | 13 37 57.3 | -09 30 41 | | | 11.220 | .002 | 10.868 | .009 | |
| 3 | 13 38 03.6 | -09 37 28 | | | 16.801 | .015 | 15.762 | .012 | |
| 3 | 13 38 07.2 | -09 30 40 | | | 13.693 | .004 | 13.256 | .012 | |
| 3 | 13 38 08.5 | -09 39 48 | | | 13.641 | .007 | 13.308 | .006 | |
| 3 | 13 38 13.3 | -09 34 10 | | | 15.331 | .008 | 15.102 | .006 | variable? |
| | | | | | | | | | |
| 4 | 13 39 28.7 | -09 58 34 | 16.293 | .009 | 15.276 | .007 | 14.942 | .008 | |
| 4 | 13 39 35.3 | -09 57 47 | 13.496 | .002 | 12.463 | .004 | 12.123 | .003 | |
| 4 | 13 39 37.7 | -09 57 17 | 15.822 | .007 | 15.127 | .006 | 14.879 | .007 | |
| 4 | 13 39 38.7 | -09 54 17 | 18.976 | .038 | 16.432 | .020 | 15.491 | .017 | |
| 4 | 13 39 40.7 | -09 58 35 | 18.727 | .029 | 16.471 | .011 | 15.599 | .009 | |
| 4 | 13 39 44.0 | -10 00 34 | 17.017 | .010 | 14.578 | .006 | 13.665 | .005 | |
| 4 | 13 39 50.2 | -09 59 53 | 15.894 | .009 | 13.510 | .004 | 12.641 | .004 | |
| 4 | 13 39 50.4 | -09 57 41 | 17.224 | .010 | 15.828 | .007 | 15.344 | .008 | |
| 4 | 13 39 53.2 | -09 53 50 | 17.015 | .019 | 14.718 | .015 | 14.175 | .019 | |
| 4 | 13 39 56.1 | -09 59 15 | 17.625 | .015 | 16.269 | .009 | 15.834 | .010 | |
| 4 | 13 39 57.1 | -10 02 59 | 15.559 | .011 | 14.356 | .005 | 13.988 | .005 | |
| 4 | 13 40 01.3 | -09 54 21 | 19.002 | .036 | 16.652 | .015 | 15.872 | .021 | |

Fig.1. Contours of flux levels from 900-second long slit spectrum of comet Temple 1 taken 18 minutes (mid-exposure) before impact.  The slit was aligned in the east-west direction. Contours are scaled between the minimum and maximum values for this exposure.

Fig. 2. Contours of flux levels from 900-second long slit spectrum of comet Temple 1 taken 49 minutes (mid-exposure) after impact. Contours are scaled between the minimum and maximum values for this exposure.

Fig. 3. All the spectra of the inner coma taken near the impact event.  Typical slit height extracted from the long-slit spectra was 15″.  Multiple consecutive exposures on 3 July and 5 July UT were averaged..

Fig. 4.  Plots of the Volume Scattering Function showing almost a factor of two increase in the scattering at blue wavelengths in 17 minutes.

Fig. 5.  The observed variation in total flux in the CN (0,0) band during the DI event. The scale on the left pertains to the R-band measurements and that on the right to CN.

Fig. 6. A montage of four R-band images taken 2 minutes before impact; and 21 minutes, 71 minutes and 25 hours after impact. The +25 hrs field is 43″ (27860 km) square.  The rest are 11″ (7128 km) square.  An image of the comet taken 2 minutes before impact was subtracted from the +25 hours co-added image to allow the broad dust fan to seen in greater detail. The yellow arrow points toward the Sun. The nominal position angle of the comet tail would be in the anti-Sun direction. The white arrow points in the direction of the initial velocity of the DI impact cloud.

Fig. 7. The mean velocity of ejecta grains during the first 25 hours after impact as measured in 6 R-band images of the dust fan. The position angle is reckoned eastward from north. The arrows indicate the position angles of the DI event and the Sun. The plotted trend line is a linear least squares fit to the data.

Fig. 8: Flux within a series of apertures (expressed as diameters of a cylinder at the distance of the comet) observed in B, R, and I on the nights of June 29 and July 5. The June 29 curves are the means of eight observations in R and six in I with only one observation in B. The July 5 curves are means of two observations in B, twelve in R, and ten in I. They were made in two sessions separated by about two hours. Observations in the three filters have been separated by adding $10 \times 10^{-15}$ ergs cm$^{-2}$ s$^{-1}$ Å$^{-1}$ to the blue observations and $5 \times 10^{-15}$ to the red. The errors in the R and I filters, expressed in units of the ordinate, are about 0.05 for the smallest aperture, decreasing to 0.01 for the third aperture and then increasing, proportionally to the flux to 0.06 for the largest. The errors in the B filter are, because of their smaller number, two to three times as large.

Fig. 9: Color index, B-R and R-I, as a function of radius on 29 June and on 5 July. Each point is determined from one to eight observations in two colors, as described on Fig. 8, and corresponds to an annulus whose inner and outer radii are the radii of two successive apertures. The abscissa is the mean of the inner and outer radii. For the R-I colors the mean

errors in magnitude in the four smallest annuli are 0.05 or less except for the second point which has an error of about 0.12. The possible source of this uncertainty is discussed in the text. The errors in the remaining annuli are about 0.015 magnitude. For the B-R colors the errors follow the same pattern but are generally about 1.5 times as large due to the small number of B observations.

Fig. 10. Flux within the same apertures as in Fig. 8 observed on the night of July 3 using only the R-filter. Two sets of six observations each, separated by about two hours, are plotted separately, showing that during this interval there has been only a very slight change. The mean errors in the unit of the ordinate are 0.04 for the leftmost aperture, decreasing to 0.01 for the third aperture and increasing thereafter, in proportion to the flux, to 0.05 for the largest aperture.

Fig. 11. Flux within the same apertures as in Fig. 8 observed on the night of impact, July 4. Results of observations with the R- and I-filters are shown for two sets of observations, approximately an hour before (solid lines) and an hour after (dashed lines) impact respectively. The first group contains four observations in R and three in I. The second group contains four observations in R and two in I. The errors for all four curves show the same pattern, starting with a large value for the smallest aperture, decreasing sharply to a minimum for the fourth aperture and thereafter increasing roughly in proportion to the flux to the largest aperture. For each of the four curves the values for these three apertures, expressed in units of the ordinate, were as follows: for I before impact, 0.015, 0.012, 0.12; for R before impact, 0.05, 0.01, 0.04; for I after impact, 0.13, 0.01, 0.10; for R after

impact, 0.46, 0.02, 0.19. Variations in brightness during the interval of observation, particularly after impact, may well have contributed to the scatter.

Fig. 12: Color index, R-I, as a function of radius on July 4, before and after impact. The observations used are the same as those displayed in Fig. 11 and the annuli are those described in Fig. 9. The mean errors in the before impact colors are about 0.04 for the two smallest annuli and 0.025 for the others. The mean errors in the after impact colors are much larger, due mainly to the much brighter comet nucleus combined with the curve-of-growth effects discussed in the text. The three smallest annuli have errors of 0.05, 0.25 and 0.18, respectively. The remainder have errors of about 0.04. The color spike at 1200 km is subject to the cautions expressed in the text.

Fig. 13. R-I for Mie scattering of sunlight from water ice spheres at a scattering angle of 139 degrees.

Fig.14. The light curves on the night of impact, showing flux as a function of time after impact for three selected apertures. Each point corresponds to a single observation. The I-filter observations (open symbols) have been shifted to the scale of the R-filter observations (filled symbols) as described in the text. The errors (for the 100 second exposures) are 0.05 before impact, increasing to 0.25 after impact, in the 3076 km aperture, 0.02 to 0.05 in the 5713 km aperture and 0.04 to 0.09 in the 10986 km aperture.

Fig. 15. R-band flux light curve of the impact ejecta constructed by subtracting the flux measured in each aperture just prior (-2 min.) to impact from the fluxes measured at later times. As expected, the flux in large apertures equals the flux in the small ones. In 80 minutes the cloud of ejecta is still totally within the first aperture. The errors are 0.25 in the 4.7″ aperture, 0.05 to 0.07 in the 8.7″ aperture and 0.06 to 0.11 in the 16.7″ aperture.

Fig. 16. A simple optically thick sphere expanding at constant velocity fit to the concave portion of the light curve 2 to 18 minutes after impact. The error bars are the same as in Fig. 15.

Fig. 1

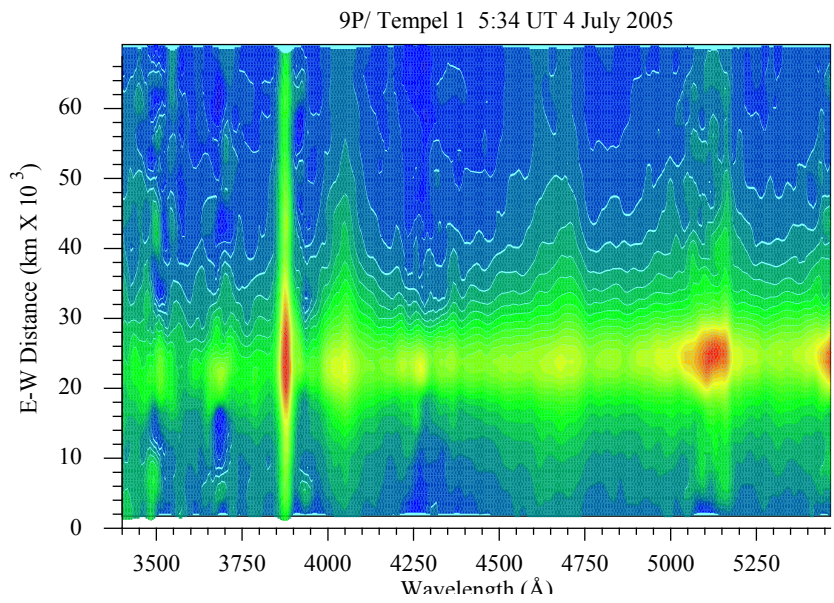

Fig. 2

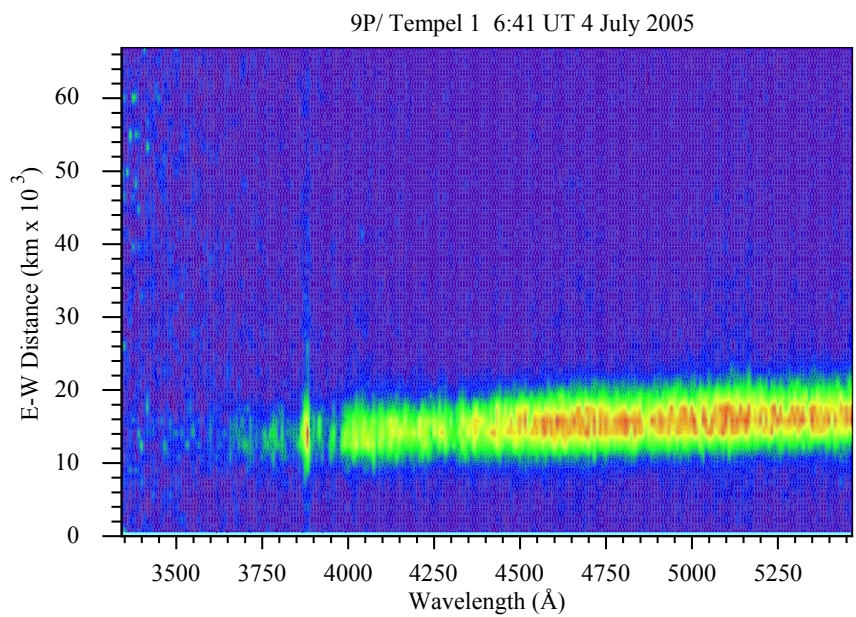

Fig. 3

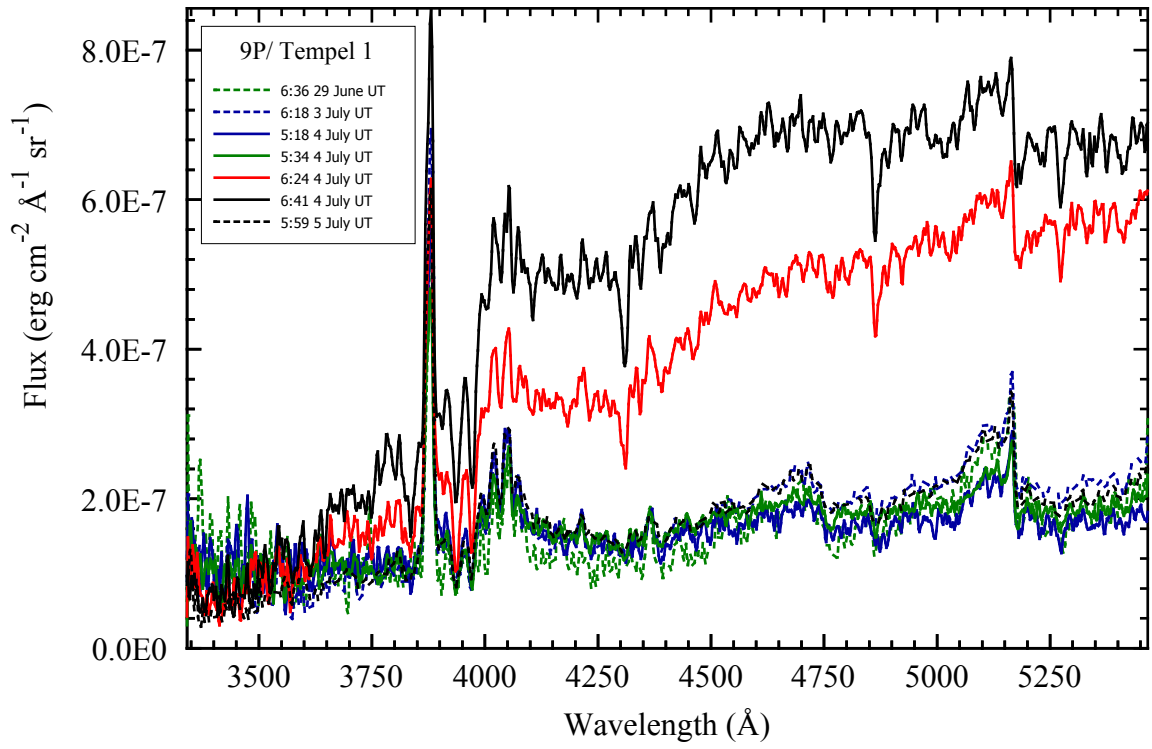

Fig.4

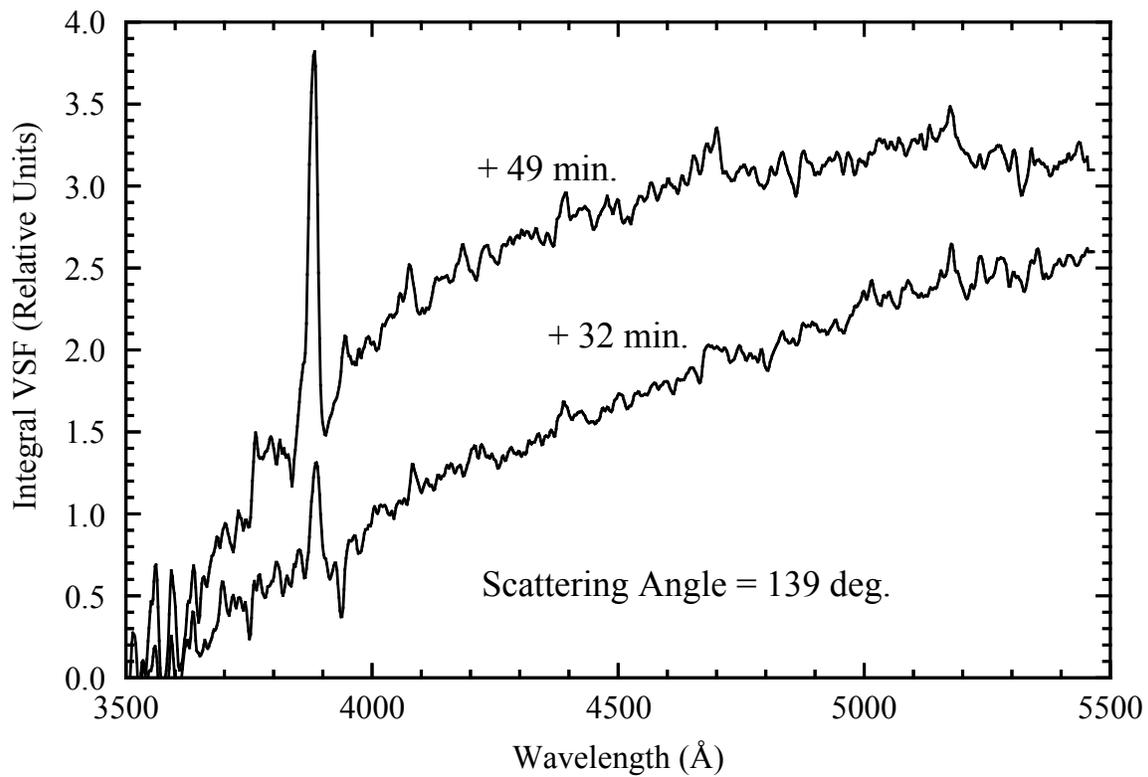

Fig. 5

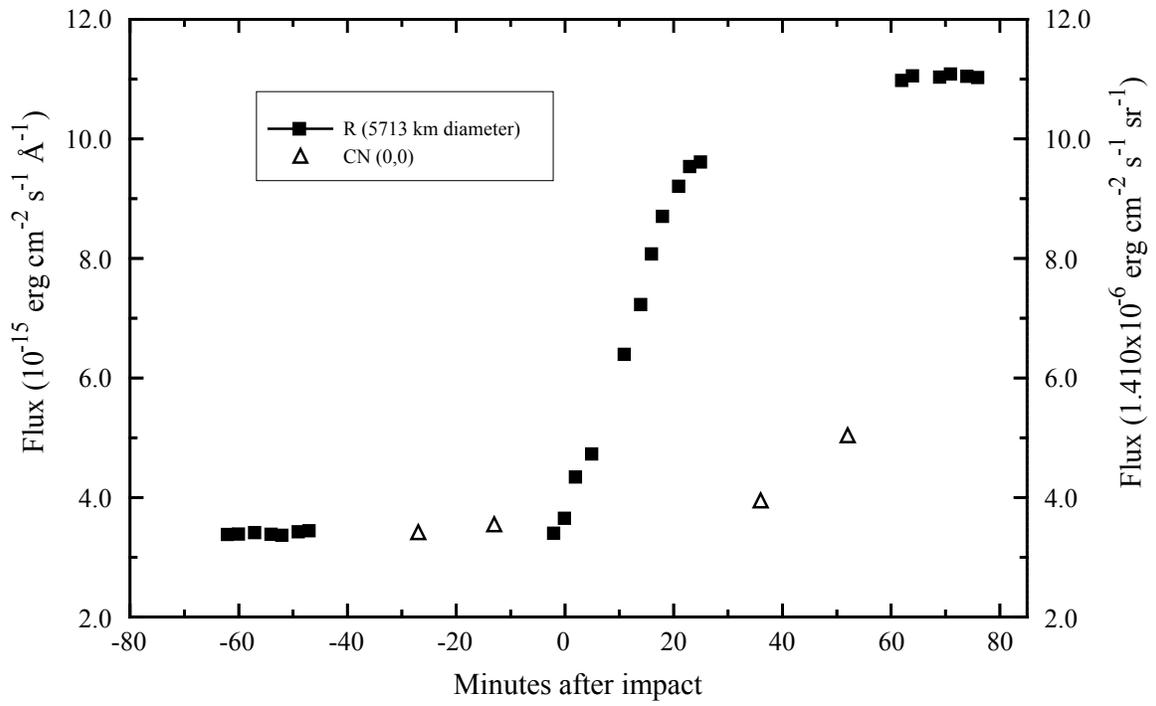

**Fig. 6**

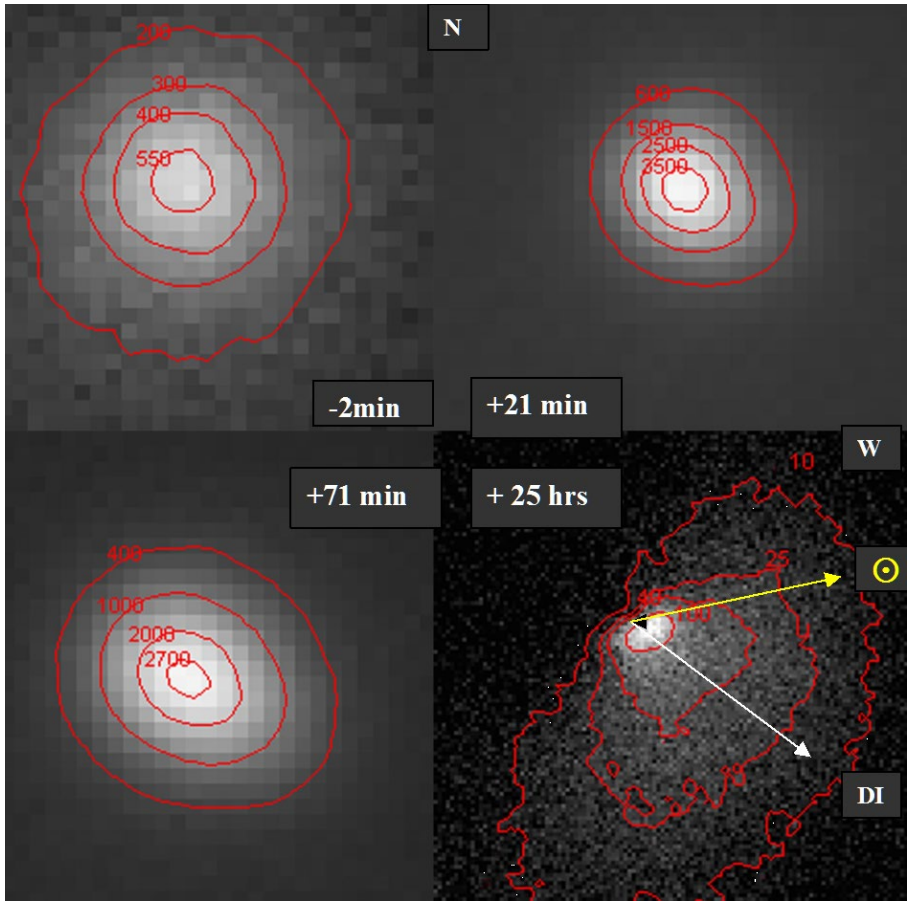

Fig. 7

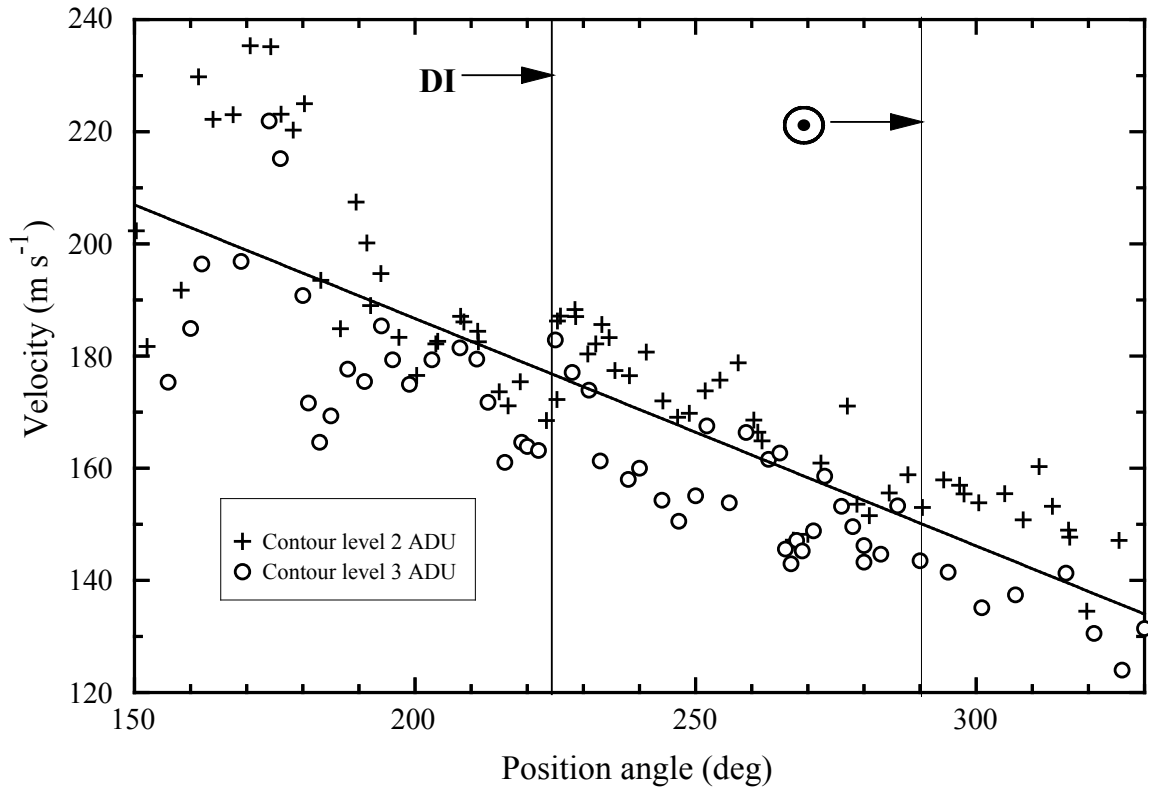

Fig. 8

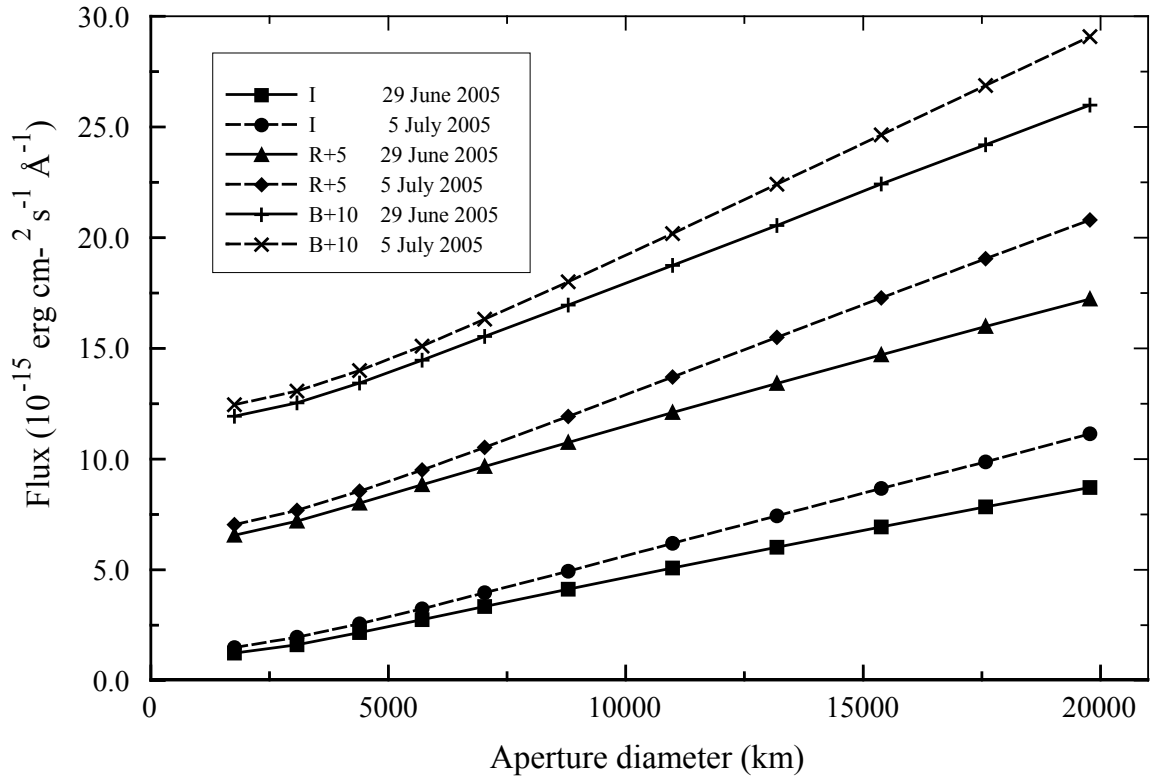

Fig. 9

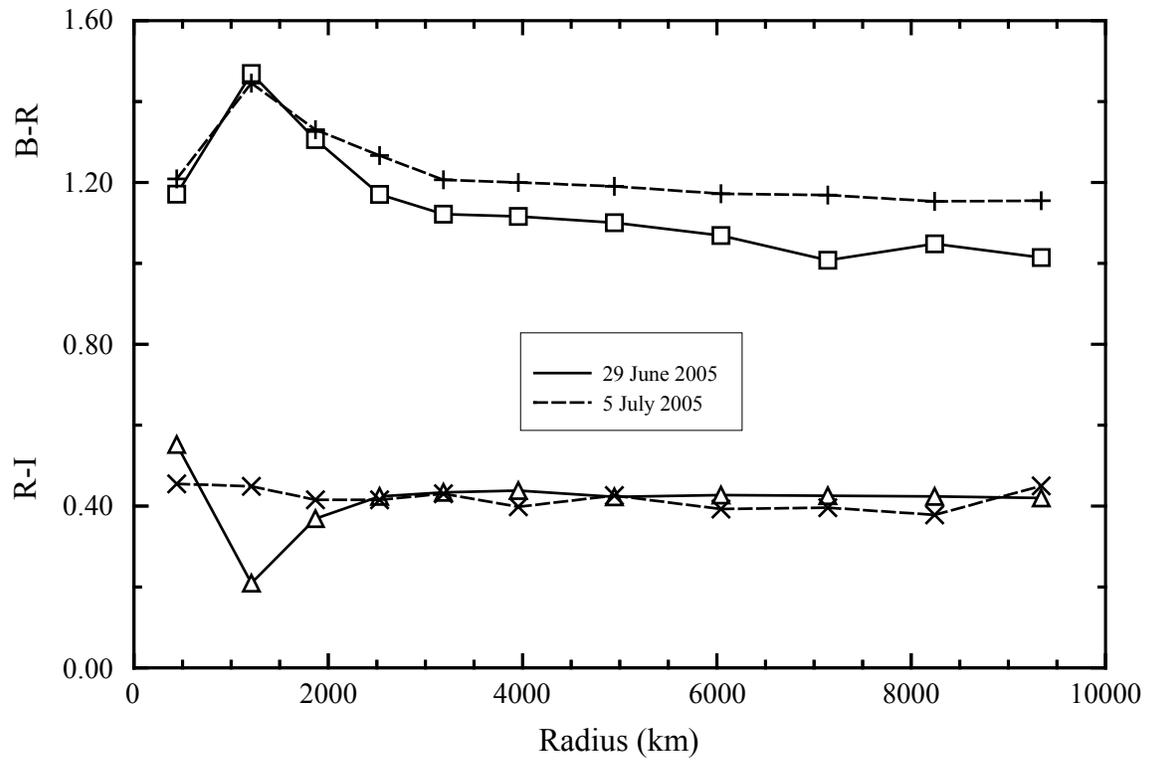

Fig. 10

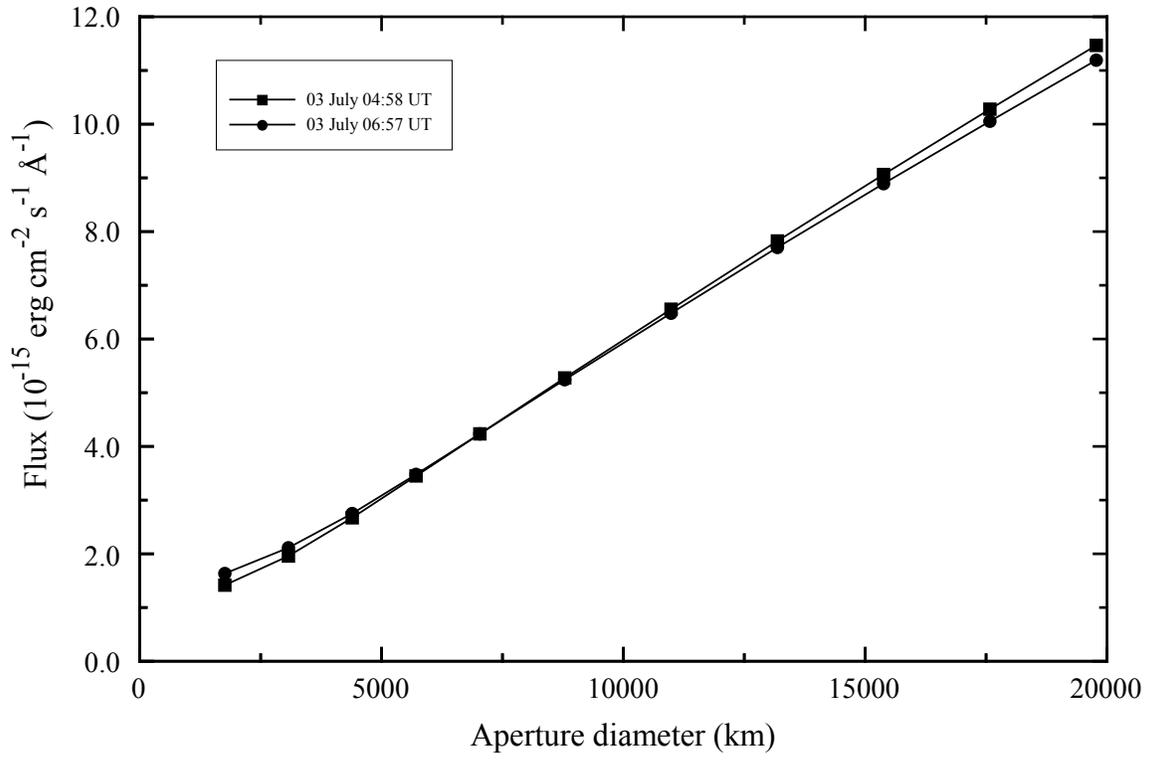

Fig. 11

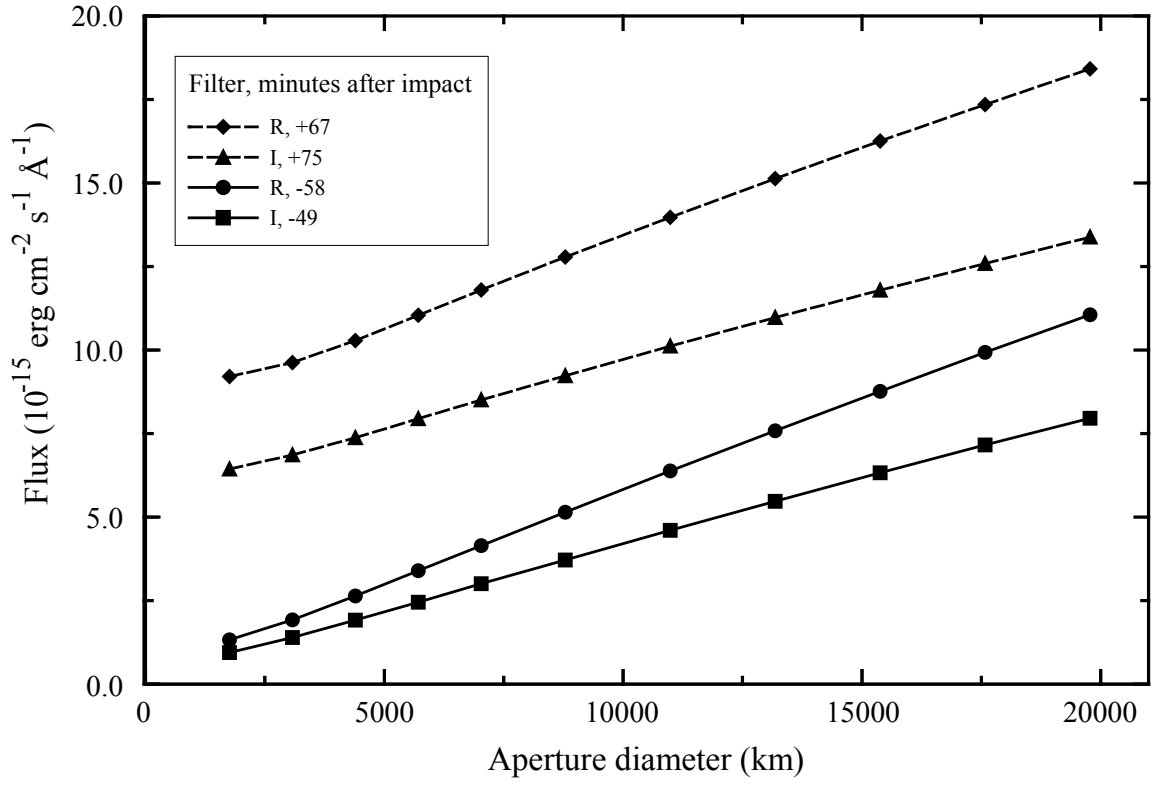

Fig. 12

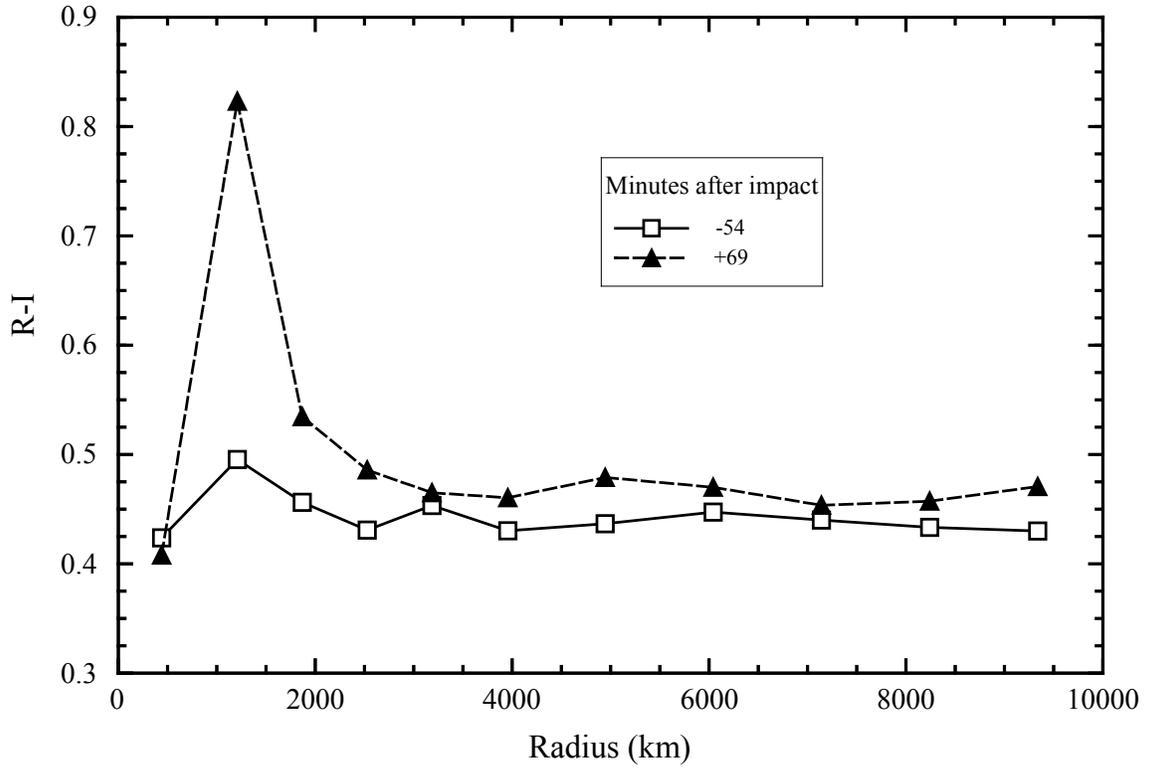

Fig. 13

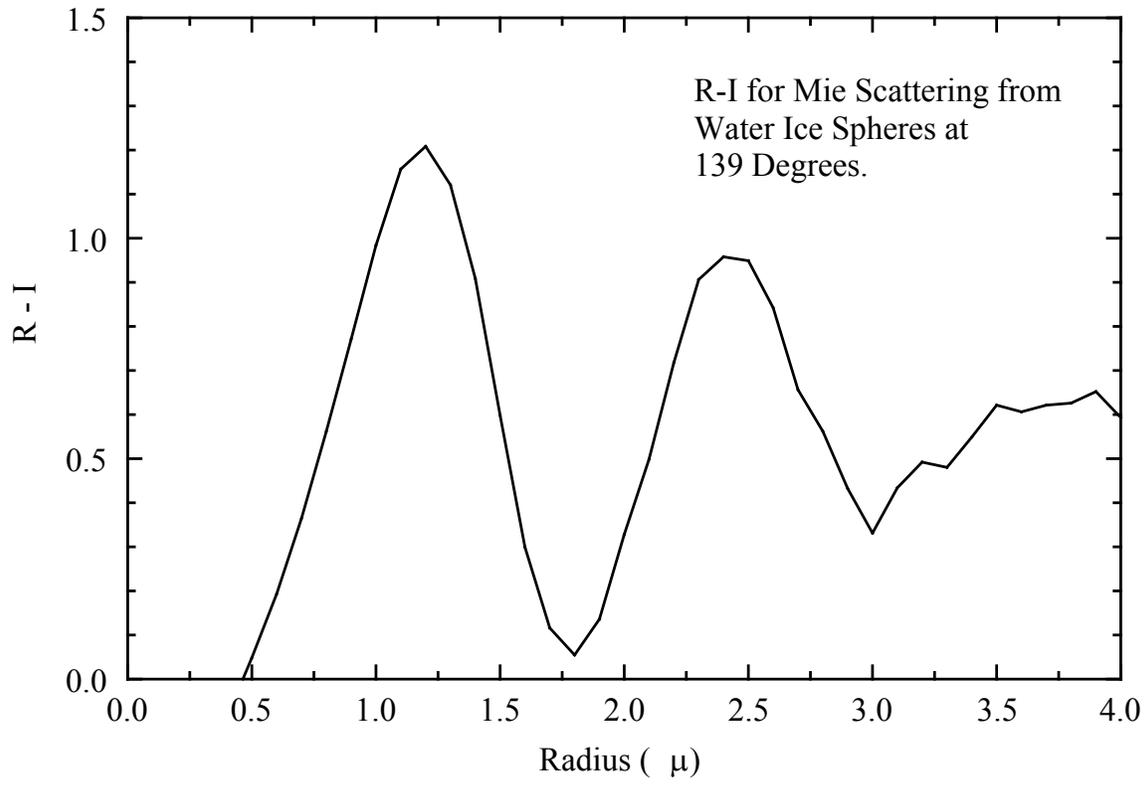

R-I for Mie Scattering from Water Ice Spheres at 139 Degrees.

Fig. 14

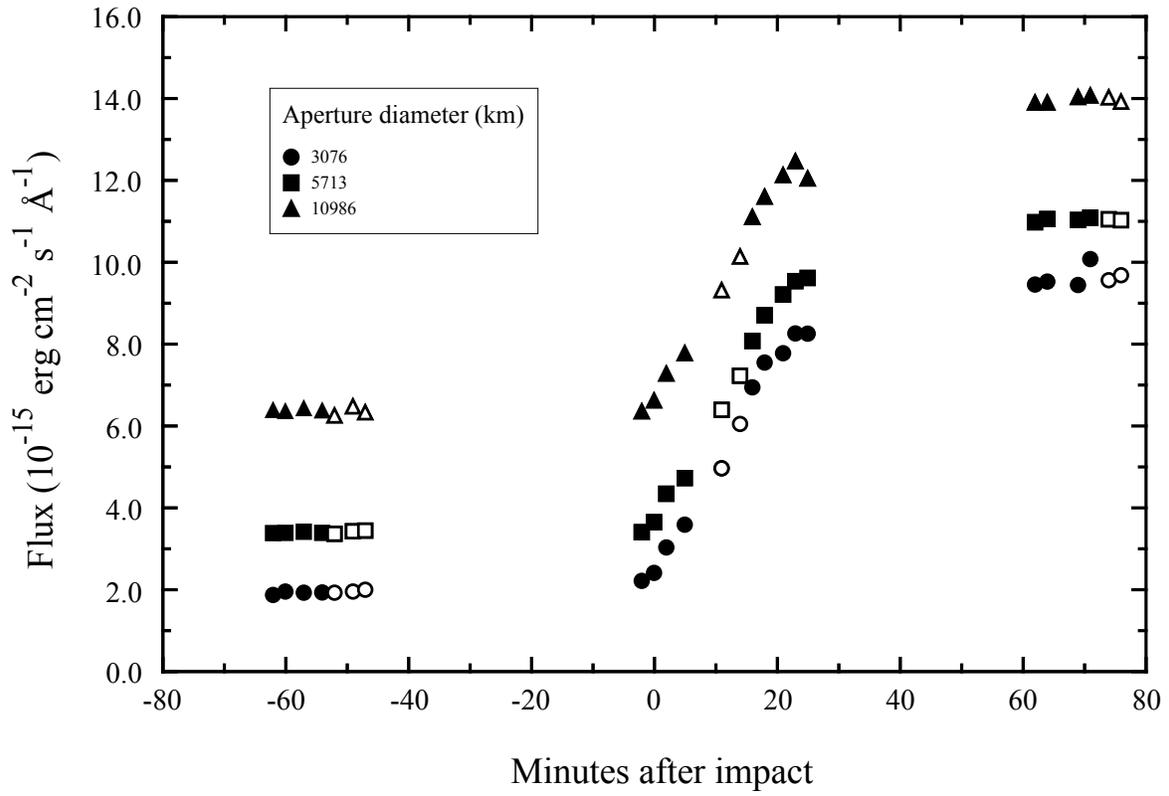

Fig. 15

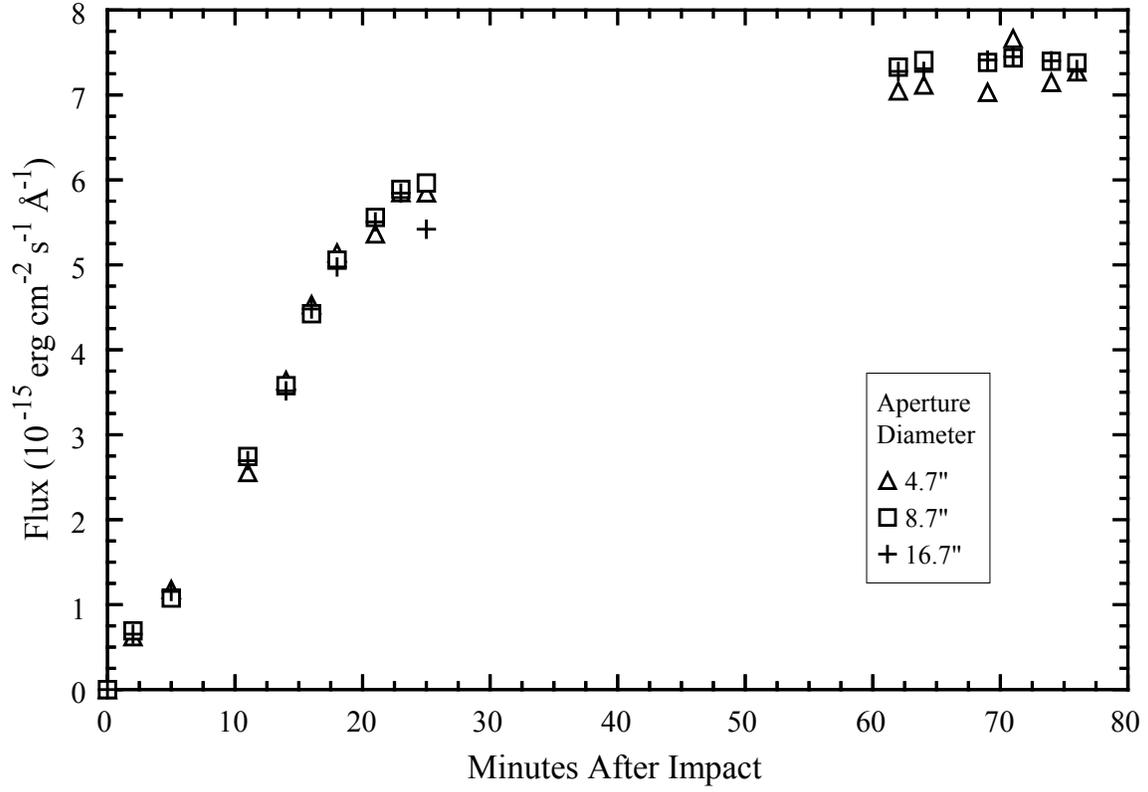

**Fig. 16**

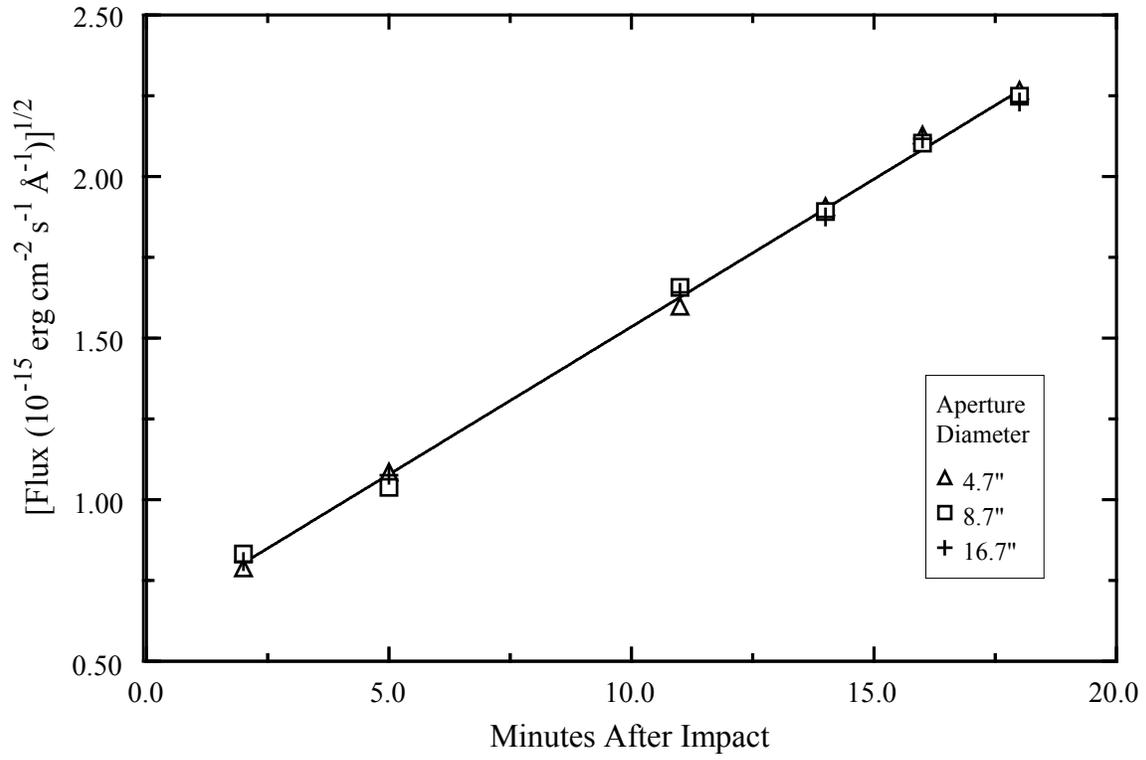